\documentclass[pra,twocolumn,showpacs,superscriptaddress]{revtex4-1}
\usepackage{color,verbatim,graphics}

\usepackage{graphicx,times}
\usepackage{amsmath,amsthm,amssymb}

\newcommand{\ME}{|G|'}

\begin{document}

\title{Necessary detection efficiencies for secure quantum key distribution and bound randomness}
\author{Antonio~Ac\'in}
\affiliation{ICFO-Institut de Ciencies Fotoniques, Mediterranean
Technology Park, 08860 Castelldefels (Barcelona), Spain}
\affiliation{ICREA-Instituci\'o Catalana de Recerca i Estudis Avan\c cats, Lluis Companys 23, 08010 Barcelona, Spain\\
}

\author{Daniel Cavalcanti}
\affiliation{ICFO-Institut de Ciencies Fotoniques, Mediterranean
Technology Park, 08860 Castelldefels (Barcelona), Spain}

\author{Elsa Passaro}
\affiliation{ICFO-Institut de Ciencies Fotoniques, Mediterranean
Technology Park, 08860 Castelldefels (Barcelona), Spain}

\author{Stefano Pironio}
\affiliation{Laboratoire d'Information Quantique, Universit\'e Libre de Bruxelles (ULB), Bruxelles, Belgium}

\author{Paul Skrzypczyk}
\affiliation{ICFO-Institut de Ciencies Fotoniques, Mediterranean
Technology Park, 08860 Castelldefels (Barcelona), Spain}
\affiliation{H.~H.~Wills Physics Laboratory,~University of Bristol,Tyndall Avenue, Bristol, BS8 1TL, United Kingdom.}

\date{\today}

\begin{abstract}
In recent years, several hacking attacks have broken the security
of quantum cryptography implementations by exploiting the presence
of losses and the ability of the eavesdropper to tune detection
efficiencies. We present a simple attack of this form that applies
to any protocol in which the key is constructed from the results
of untrusted measurements performed on particles coming from an
insecure source or channel. Because of its generality, the attack
applies to a large class of protocols, from standard
prepare-and-measure to device-independent schemes. Our
attack gives bounds on the critical detection efficiencies
necessary for secure quantum distribution, which show that the
implementation of most partly device independent solutions is,
from the point of view of detection efficiency, almost as
demanding as fully device-independent ones. We also show how our
attack implies the existence of a form of bound randomness, namely
non-local correlations in which a non-signalling eavesdropper can
find out a posteriori the result of any implemented measurement.
\end{abstract}

\maketitle

Over the past few decades the problem of bridging the gap between
realistic implementation of Quantum Key Distribution (QKD)
protocols and their theoretical security proofs has attracted a
lot of attention. The security of standard QKD protocols
\cite{BB84,E91} relies on a very detailed modeling of the
preparing and measuring devices. However, unavoidable
imperfections of the devices or unnoticed failures lead in
practice to deviations from the model used to prove security --
deviations that can be taken advantage of by a potential
eavesdropper. Indeed, standard QKD protocols, being dependent on
the accuracy with which the devices are described, can typically
suffer attacks, for instance on the detectors~\cite{LWWESM10}.

To overcome these problems a new paradigm was proposed, adopting
the device-independent (DI) framework \cite{DIQKD}: In this scenario no assumptions
are made either on the source of the shared system or on the
internal working of the devices, which are treated like ``black boxes''. In this context
the only object one relies on is the statistics of inputs and
outputs, and the security of a device-independent quantum key
distribution (DIQKD) protocol is guaranteed by the nonlocal
character of these statistics \cite{NL review}. The DI scenario
allows for the most general and powerful quantum certification
protocols as     it depends on very few assumptions. Nevertheless,
their implementations are demanding because they
require very high detection efficiencies to close the detection
loophole (e.g. with photonic implementations~\cite{NL review,bellexp}).

In order to make the experimental implementations less demanding other
scenarios between standard and fully DI QKD have
been introduced. In these intermediate scenarios the parties
involved add some extra assumptions to the
fully-DI scheme. The focus is still on the
input/output statistics but with an intermediate level of trust
between the fully-DI framework and the
device-dependent one. For instance, semi-device-independent (SDI)
protocols have been proposed where one makes an assumption on the
dimension of the involved quantum systems but, apart from this
assumption, the devices are still uncharacterized~\cite{PB11}.
From an implementation point of view, the advantage of SDI
protocols is that they do not require entanglement and can be
implemented in a prepare-and-measure configuration. Another class
of intermediate scenario, known as one-sided device-independent
(1SDI) \cite{TR11,TLGR12,BCWSW12}, is based upon quantum steering \cite{WJD07} which consists of a bipartite scenario in which one of the parties trusts his measuring devices but the other does not.

All these different QKD solutions are based on different
assumptions and, thus, offer different levels of security. Although different QKD protocols use different strategies,
most of them share the property that the key is constructed from the results of measurements performed by one of the end-users on quantum particles that have propagated
through an insecure channel. This is the case, for instance, of the famous Bennett-Brassard-84~\cite{BB84} and
Ekert~\cite{E91} protocols, and standard DIQKD protocols, such as those introduced in~\cite{amp,DIQKD}.
Notice however that not every QKD protocol is
of this form, a paradigmatic example being
measurement-device-independent QKD~\cite{mdiqkd,mdiqkd2}.

In this work, we consider the above scenario and therefore focus
on an end-user in a cryptographic protocol who performs
measurements on some quantum systems received through an insecure
channel. We introduce an attack by an eavesdropper who is
able to control the detection efficiency of the measurements -- a
natural assumption in the adversary model of cryptographic
protocols based on untrusted measurements, such as 1SDI, SDI, and
DI protocols. The attack also applies to standard
prepare-and-measure protocols if one cannot guarantee that the
eavesdropper is unable to tune the detection efficiencies. In fact
recent hacking attacks on standard QKD protocols have exploited
the ability to manipulate detection efficiencies \cite{LWWESM10}.
Our attack defines detection efficiencies necessary for secure
quantum key distribution using the previous protocols. We then discuss how our attack can also be applied to schemes for
randomness generation. From a practical point of view,
our results imply that the implementation of partly DI protocols
are, in terms of detection efficiency, almost as demanding as
fully DI ones. Moreover, our attack has also implications from a
fundamental point of view: as also observed independently in
\cite{Eric1,Eric2}, it implies the existence of a very weak form
of intrinsic randomness in which an eavesdropper limited only by
the no-signalling principle~\cite{BHK05} cannot a priori fix the
outputs of the measurements in a Bell test, but she can later find
out the result of any implemented measurement. In analogy with
results in thermodynamics and entanglement theory~\cite{BE} we
name this effect \emph{bound randomness}.

\section{The attack} The considered scenario consists of a party, say
Bob, who measures quantum systems received through an insecure
channel. The received systems may have been prepared by another
honest party, say Alice, or by an untrusted source. In
particular, they may be entangled with other quantum systems.
Bob performs on them one of $M_B$ possible measurements with $D$
possible outcomes. We label the measurement choice and result by
$y=1,\ldots,M_B$ and $b=1,\ldots,D$ respectively. In the absence of loss, let Bob's device give the outcome $b$ with probability $Q(b|y,\rho)$, where $\rho$ is the state of the system received by Bob and which may be correlated with classical or quantum variables of other parties in the protocol. For simplicity in the notation, we omit $\rho$ in what follows, as our results are independent of it.

In a realistic implementation with losses and inefficient detectors, each measurement of Bob will have a detection
efficiency $\eta_y$, and one more outcome is observed, corresponding to
the no-click events which we denote by $b=\emptyset$. That different measurements may have different efficiencies naturally arises in certain situations, e.g. in \cite{largeviol}. In such a situation, Bob's device then produces outcomes with probabilities $P(b|y)=\eta_y Q(b|y)$ for $b = 1,\ldots, D$, and
$P(\emptyset|y)=1-\eta_y$.

We exhibit here below a simple attack which allows Eve to learn the
output of any subset $G\in\{1,\ldots,M_B\}$ of Bob's measurements. This
attack does not modify any of Bob's outcome probabilities, i.e., it
reproduces the full lossy behavior of Bob's device. In particular, we
stress that it does not rely on Bob performing any kind of
post-selection. The attack requires that Eve is able to tune arbitrarily the detection efficiency of Bob's detectors depending on the
implemented measurement and works as long as Bob's observed detector efficiencies satisfy $\sum_{y\in G}\eta_y \leq 1-\eta'$, where $\eta'=\max_{y\notin G}\eta_y$ is the maximum detection efficiency over the set of measurements complementary to $G$, i.e., those that Eve is not interested in guessing. (If this complementary set of measurements is empty, i.e. when Eve wants to guess the output of all of Bob's measurement, we define $\eta'=0$).

In the simple case where all detectors have the same efficiency $\eta_y = \eta$, the attack works whenever $\eta\leq 1/(|G|+1)$ if $|G|<M_B$ or when $\eta\leq 1/M_B$ if $|G|=M_B$. In particular, when Eve is interested in guessing a single one of Bob's measurements, say $\bar y$, then $|G|=1$ and the attack works as long as $\eta \leq 1/2$. Furthermore, if the detectors are not all equally efficient, Eve can use the inefficiency of the measurements $y\neq \bar y$ that she is not interested in to raise the critical efficiency of the measurement $\bar y$ that she wants to guess above $\eta_{\bar y}=1/2$, as long as $\eta_{\bar y}\leq 1-\max_{y\neq \bar y}\eta_y$.

Let us now explain how the attack works. Eve randomly selects with probability $\eta_y$ one of the measurement $y\in G$ whose outcomes she wants to guess and with probability $1-\sum_{y\in G}\eta_y$ she does not select any particular measurement. Depending on her choice, she then applies one of the two following strategies.\\
$(i)$ If she picked measurements $\bar y\in G$, she performs this measurement on the incoming state. She obtains outcome $b$ with probability $Q(b|\bar y)$, she reads the outcome, and forwards the corresponding reduced state to Bob. On Bob's side, she forces Bob's detector to click if he performs measurement $y=\bar y$, in which case he obtains the same outcome $b$. If otherwise $y\neq \bar y$, she instructs Bob's device not to click, i.e., to output $b=\emptyset$.\\
$(ii)$ If she didn't select any particular measurement, she directly forwards the state to Bob without intervention. However, she instructs Bob's device not to click ($b = \emptyset$) if $y\in G$. If on the other hand $y \not\in G$, she allows his detector to click with probability $\tau_y$. Bob then obtains a proper result $b$ with probability $\tau_y Q(b|y)$ and a no-click result with probability $1-\tau_y$.

Obviously, Eve can always correctly guess Bob's output when $y\in G$ since when Bob's measuring device clicks, it always coincides with Eve's previous measurement result, and she always knows when his detector does not click (gives outcome $b = \emptyset)$. Moreover, defining the $\tau_y$ such that $\eta_y=(1-\sum_{y\in G}\eta_y)\tau_y$ for $y\notin G$, it is straightforward that the strategy yields the overall outcome probabilities $P(b|y)=\eta_y Q(b|y)$ if $b\neq \emptyset$ and $P(\emptyset|y)=1-\eta_y$, which correspond to lossy devices characterized by detection efficiencies $\eta_y$. The only requirement for the $\tau_y$s to be well-defined is that $\sum_{y\in G}\eta_y\leq 1-\eta'$, where $\eta'=\max_{y\notin G}\eta_y$.

\subsection{Application to QKD protocols} The above attack
applies to any cryptographic protocol in which the key is constructed from the results of measurements performed by one of the end-users on quantum particles
received through an insecure channel. It thus applies to any Bell
based DI protocol, but also to SDI approaches where the dimension is fixed, protocols
based on steering, or prepare-and-measure protocols, unless the eavesdropper cannot tune Bob's detection efficiencies. In fact, in many of these protocols, the key is
constructed from a single measurement, which means that in the best case scenario (that of equal detection efficiencies) they
become insecure at $\eta=1/2$. It is important to notice that
the obtained critical detection efficiencies apply to any
scenario, independently of the number of measurements $M_B$,
outputs $D$, or the role of other parties in the protocol.

By using many measurements for the key generation, one increases the number of measurements that Eve needs to guess and the
critical detection efficiency for our attack decreases. However,
this solution is demanding from Alice's and Bob's point of view as
many more symbols are sacrificed after basis reconciliation, and
also more statistics needs to be collected to have a proper
estimation of the protocol parameters. In fact the advantage of
using more measurements is limited when considering two distant
parties connected by a lossy channel. Take for instance a rather
idealised situation in which all losses come from the channel,
denoted by $\eta_C$ and are equal to $\eta_C=10^\frac{-\alpha L}{10}$
where $L$ is the distance in km. Then, the improvement in distance
with the number of bases is only logarithmic.
For instance, assuming a typical value for the losses of $\alpha$
of the order of $0.2$ dB/km, one has that in order to compensate
for the channel losses at 100 km Alice and Bob need to employ 100
bases.

A possible solution to overcome channel losses is to use heralded
schemes \cite{gps,mba} or quantum repeaters based on entanglement swapping \cite{Wein}. Using such schemes, which are technologically more demanding, the only relevant losses for security
are those on the honest parties' labs. Alice and Bob can then decide which cryptographic solution to adopt,
from standard to fully device-independent, depending on the observed detection inefficiencies
and the plausibility of the assumptions needed for security.

Our attack also applies to randomness generation schemes based on
correlations between measurements on two different devices. In
these schemes, randomness is certified by the observed quantumness
of the correlations, certified for instance by means of
steering~\cite{singapore,inprep} or Bell
inequalities~\cite{colbeck,nature}. As the particles come from a
untrusted source, one cannot exclude that the attack has been implemented on each of the particles sent to the
untrusted parties in the protocol (one in the case of steering and two
for Bell-based schemes).

In the case of Bell-based protocols, for instance, it is
possible to guess the result of one measurement on each device when their
detection efficiency is 1/2. Note that in the context of randomness expansion,
it is usually the case that one of the possible combinations of measurements
is implemented most of the time, as this
requires much less initial randomness to run the Bell
test~\cite{nature}. For all these protocols, randomness
expansion is lost when the critical detection efficiency
is $1/2$.

\subsection{Improved attacks} The previous attack applies to many
cryptographic scenarios because it is independent of the number of
measurements, outputs and actions by other parties. Improvements
however may be expected for concrete protocols. For instance, we
show in what follows how for two untrusted measuring devices,
Eve can improve the attack by exploiting the detection efficiency of
the second party too. Note though that the attack needs more operations from Eve's side
on the untrusted devices than just varying the detection efficiency of the
implemented measurements. This improved attack is inspired by the
local models exploiting detection inefficiencies introduced in~\cite{mp}.

We thus consider a second party in the protocol, Alice,
who performs $M_A$ measurements of $D$ outputs. Her measurement
choice and result are labeled by $x$ and $a$. Again, in the presence of loss, the output probability distribution has one more result because of the no-click events and is of the form $P(ab|xy) = \eta^2 Q(ab|xy)$, $P(\emptyset b|xy) = \eta(1-\eta) Q(b|y)$, $P(a \emptyset|xy) = \eta(1-\eta)Q(a|x)$, $P(\emptyset \emptyset|xy) = (1-\eta)^2$,
where the detection efficiencies have for simplicity all been taken to be equal to $\eta$.

In the improved attack, Eve's goal is again to guess $G$
measurements on Bob's side. With probability $q$ Eve uses the previous
attack and does nothing on Alice's side. With probability $1-q$
the attack works in the reverse direction: Eve fixes
the output of one of Alice's measurements (even though she is still guessing Bob's result). That is, she picks one of Alice's
measurements, say $\bar x$, with probability $1/M_A$, and decides an output
for this measurement following the quantum probability $Q(a|\bar x)$.
If Alice happens to implement measurement $\bar x$ she will obtain this outcome, otherwise
she observes a no-click. On Bob's side, Eve computes the reduced state corresponding to
Alice's result and, for each measurement by Bob, selects one
possible outcome following the probability $Q(b|y,ax)$ predicted by this state. This defines Bob's
result, whose detector always clicks. The
intuition behind the attack is that for those cases in which Eve
fixes Alice's result, she can allow any measurement on Bob to give
a result, as Alice effectively implements one single measurement
and a hidden-variable model is enough to describe the observed correlations.

So far the model never gives two
no-click events, which does not correspond to the expected behavior of actual lossy devices. To correct this, with probability $r$, Eve runs the above protocol and with
probability $1-r$, she instructs both detectors not to click. We finally get
\begin{equation}
\begin{aligned}
P(ab|xy) &= r\left( \frac{q}{\ME} +  \frac{1-q}{M_A}\right) Q(ab|xy)\\
P(a \emptyset|xy) &= r\,q \left(1-\frac{1}{\ME}\right)Q(a|x)\\
P(\emptyset b|xy) &= r(1-q)\left(1-\frac{1}{M_A}\right)Q(b|y)\\
P(\emptyset \emptyset|xy) &= 1-r=(1-\eta)^2  ,
\end{aligned}
\end{equation}
where $\ME=|G|+1$ when $|G|<M_B$ and $\ME=|G|$ when $|G|=M_B$, as in the previous attack. Tuning the parameters so that the above probabilities correspond to those of lossy devices with equal efficiencies $\eta$, one finds
\begin{equation}
\eta=\frac{\ME+M_A-2}{\ME M_A-1} .
\end{equation}
It is easy to see that this attack improves over the previous one, as the corresponding critical detection efficiency
is always larger than $1/\ME$. For example, in the simplest case where Alice performs 3 measurements, Bob performs two, and Eve guesses a single outcome, $(M_A,M_B,|G|) = (3,2,1)$, $\eta = 3/5$, increasing the critical efficiency by a further $10\%$. In the opposite limit, when $M_A\rightarrow\infty$, $\eta \rightarrow 1/\ME$, showing that
 the advantage of attacking Alice's measurements decreases with the number of measurements she performs.

\section{Bound randomness} Our results are not only limited to practical aspects of
cryptographic protocol implementations, but also have implications
from a more fundamental point of view. Indeed, they imply the existence (see also \cite{Eric1, Eric2}) of
non-local correlations with a very weak form of randomness
in which an eavesdropper (i) cannot obviously fix
the results of all measurements in advance but
(ii) can later predict with certainty the outcome of any measurement.
As mentioned, we dub this effect bound randomness. Our last result
is to show the existence of bound randomness in the case of eavesdroppers limited only
by the no-signalling principle~\cite{BHK05}.

The construction of bound randomness relies on a couple of simple
observations. First, in a randomness scenario consisting of
two untrusted devices with uniform detection efficiency $\eta=1/2$, our (primary) attack can be applied to both parties, so that
the eavesdropper learns the result of one measurement each
for Alice and Bob, $\bar x$ and $\bar y$. Let $e=(e_a,e_b)$ be Eve's prediction for Alice and
Bob's outcomes for measurements $\bar x$ and $\bar y$.
This variable can take $(D+1)^2$ possible values corresponding to
the ideal $D$-valued measurement outcomes plus the no-detection
event. Eve obtains outcome $e$ with a certain probability $P_{\bar x\bar y}(e)$ and given $e$, her attack defines a joint probability $P_{\bar x\bar
y}(ab|xy,e)$ for Alice and Bob. Since the attack does not change the expected probabilities $P(ab|xy)$ from Alice and Bob's perspective, we have that
\begin{equation}
\sum_e P_{\bar x\bar y}(abe| x
 y) = P(ab|xy) ,
\end{equation}
where we have defined the tripartite
conditional probability distribution $P_{\bar x\bar y}(abe| x
 y)=P_{\bar x\bar y}(e)P_{\bar x\bar
y}(ab|xy,e)$.
Now, the $M_A M_B$ different attacks
defined by each combination of measurement settings $z=(\bar x,\bar y)$ can be combined into a single
tripartite conditional probability distribution
\begin{equation}
\label{steeringcorr}
P(abe|xyz)\equiv P_z(abe|xy)
\end{equation}
by adding an input $z$ on Eve's, where $z$ defines the combination of settings
Eve wants to predict.
It is easily verified that this tripartite distribution is no-signalling, see also~\cite{renner}, and thus represents a valid attack by a no-signalling eavesdropper. By choosing her input $z$, Eve can steer the ensemble of non-signalling
correlations prepared between Alice and Bob. Thus, she can choose a posteriori the attack
that allows her to predict the result of any given pair $z$ of implemented measurements.
The effect is similar to what happens in the quantum case when
predicting the result of non-commuting variables on half of a
maximally entangled state.

Note now that there exist correlations that are non-local -- hence
whose outcomes cannot all be fixed in advance -- even when the
detection efficiency is smaller than $1/2$ -- hence whose outcomes
can all be perfectly guessed by Eve a posteriori using the above
construction. Examples of such correlations were given
in~\cite{M02}, where it was shown that the critical detection
efficiency required to close the detection loophole decreases
exponentially with the dimension of the measured quantum state in
a scenario in which the number of measurements by Alice and Bob is
exponentially large. More generally, any non-local correlations
obtained for detection efficiencies $\eta \leq 1/2$ constitute
examples of bound randomness. Finally, it can be explicitly
checked that both the all-versus nothing example of \cite{Cab} and
the Peres-Mermin magic square \cite{PeresMermin} exhibit bound
randomness \cite{footnote}.

\section{Conclusions} We have provided a simple and general detection attack that
allows an eavesdropper to guess some of (or all) the measurement results
in a cryptographic protocol.  It applies basically to any protocol
with untrusted detectors in which she is able to tune the detection
efficiency of untrusted devices. Obviously our attack cannot be
applied to protocols in which the key is not constructed from measurement
results, such as in measurement-device-independent schemes~\cite{mdiqkd,mdiqkd2}.
These protocols, almost by definition, are only sensitive to attacks on the devices that
prepare the quantum states.
The generality of our attack also implies that the implementation of
partly DI solutions is, from the point of view of detection efficiency, almost as demanding
as DI ones, which, in turn, offer stronger security.

Interestingly, the critical detection efficiency corresponding to our attack only depends on the number of measurements that Eve wants to learn, but is independent of the total number of measurements $M_B$, number of outputs $D$, or dimension of the quantum systems used.

We have also presented an improved attack that applies to protocols with
two untrusted detectors. In this attack, the eavesdropper exploits the
detection inefficiencies of one of the parties to improve her attack on
the other party. More generally, it would be interesting to
derive a formalism to study the robustness of concrete protocols
to detection attacks, as these are the most advanced at the moment.
This will allow us to understand for which protocols the detection bounds
for security derived here are tight. An analysis of the tightness of our attack in steering scenarios will be presented in \cite{inprep}.

Finally, our results imply also the existence of a bound randomness,
an intriguing and weak form of certified randomness. In a scenario in which an eavesdropper
is limited only by the no-signalling principle, there exist non-local correlations for which
she can find out a posteriori the results of any implemented measurements.
A final open question is to understand if this form of randomness exists
in the quantum case, that is, when the eavesdropper is limited by
the quantum formalism.

\begin{acknowledgements}
{Acknowledgments.} We thank M. Paw\l{}owski for useful discussions. This work was supported by the Spanish project FOQUS, the Generalitat de Catalunya (SGR 875), the ERC CoG QITBOX and the John Templeton Foundation, the EU project QALGO, by the F.R.S.-FNRS under the project DIQIP, the Brussels-Capital Region through a BB2B grant and the Interuniversity Attraction Poles program of the Belgian Science Policy Office under the grant IAP P7-35 photonics@be. DC is supported by the Beatriu de Pin\'os fellowship (BP-DGR 2013) and PS partially by the Marie Curie COFUND action through the ICFOnest program and the ERC AdG NLST. SP is a Research Associate of the Fonds de la Recherche Scientifique F.R.S.-FNRS (Belgium).
\end{acknowledgements}

\end{document}